\documentclass[a4paper,floatfix,rmp,twocolumn,showkeys,superscriptaddress]{revtex4}
  \usepackage[latin1]{inputenc}
  \usepackage{graphicx}
  \usepackage{mathtools}
  \setcitestyle{numbers,square,comma}

\begin{document}

  \title{How Turing parasites expand the computational landscape of digital life}

  \providecommand{\ICREA}{ICREA-Complex Systems Lab, Universitat Pompeu Fabra (GRIB), Dr Aiguader 80, 08003 Barcelona, Spain. }
  \providecommand{\IBE}{Institut de Biologia Evolutiva, CSIC-UPF, Pg Maritim de la Barceloneta 37, 08003 Barcelona, Spain. }
  \providecommand{\SFI}{Santa Fe Institute, 1399 Hyde Park Road, Santa Fe NM 87501, USA. }
  \providecommand{\GISC}{Grupo Interdisciplinar de Sistemas Complejos (GISC), Madrid, Spain. }
  \providecommand{\CNB}{Departamento de Biolog\'ia de Sistemas, Centro Nacional de Biotecnolog\'ia (CSIC), C/ Darwin 3, 28049 Madrid, Spain. }

  \author{Lu\'is F Seoane}
    \affiliation{\CNB}
    \affiliation{\GISC}
    
  \author{Ricard Sol\'e}
    \affiliation{\ICREA}
    \affiliation{\IBE}
    \affiliation{\SFI}

  \vspace{0.4 cm}
  \begin{abstract}
    \vspace{0.2 cm}

    Why are living systems complex? Why does the biosphere contain living beings with complexity features beyond those of the simplest replicators? What kind of evolutionary pressures result in more complex life forms? These are key questions that pervade the problem of how complexity arises in evolution. One particular way of tackling this is grounded in an algorithmic description of life: living organisms can be seen as systems that extract and process information from their surroundings in order to reduce uncertainty. Here we take this computational approach using a simple bit string model of coevolving agents and their parasites. While agents try to predict their worlds, parasites do the same with their hosts. The result of this process is that, in order to escape their parasites, the host agents expand their computational complexity despite the cost of maintaining it. This, in turn, is followed by increasingly complex parasitic counterparts. Such arms races display several qualitative phases, from monotonous to punctuated evolution or even ecological collapse. Our minimal model illustrates the relevance of parasites in providing an active mechanism for expanding living complexity beyond simple replicators, suggesting that parasitic agents are likely to be a major evolutionary driver for biological complexity.

  \end{abstract}


\maketitle

 \section{Introduction}
    \label{sec:1}

    \begin{figure*}[]
      \begin{center}
        \includegraphics[width=15 cm]{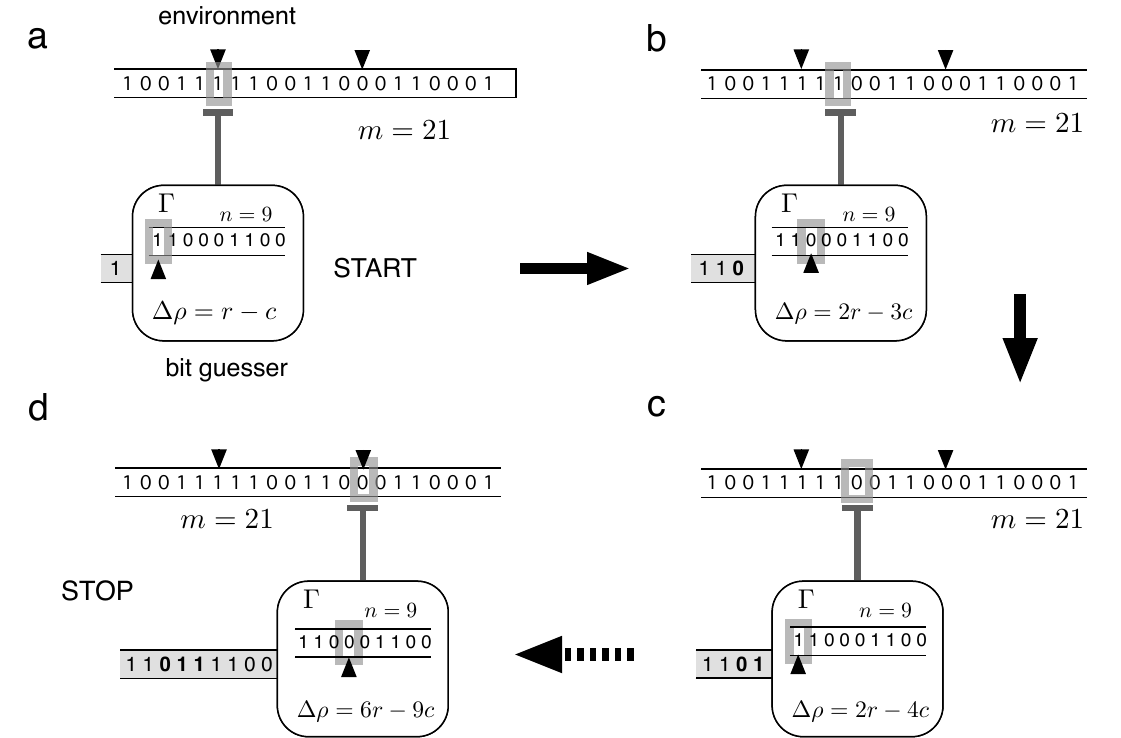}

        \caption{{\bf A bit-guesser "machine" computing its environment.} Bit-guessers use a bit-string ($\Gamma$) to attempt to predict their environment. Here $\Gamma$ approximates the most likely pattern to be found in the environment $E$---e.g.\ the first bit proposed (a) is just the most common bit in E. For each guess attempt, a cost $c$ is subtracted from guesser's reward $\rho$. For each correct guess, a reward $r$ is added. (b) After a guess, the guesser moves forward along E. If the guess is correct, the guesser keeps sampling $\Gamma$ in linear order. If the guess is wrong ((b) and (c), bits highlighted in bold), the agent resets to the beginning of $\Gamma$ for the next guess attempt. (d) The process goes on until as many bits have been visited as the size of the guesser's internal pattern $\Gamma$ (in this case, $n=9$). }

        \label{fig:1}
      \end{center}
    \end{figure*}

    With the emergence of life on Earth, information processing took on an unprecedented relevance \cite{Hopfield1994, SzathmaryMaynardSmith1997, Joyce2002, WalkerDavies2013, SmithMorowitz2016}. Swiftly, mechanisms for error correction \cite{SmithMorowitz2016}, memory (hence path dependency and contingency) \cite{Gould1990,Jablonka2006}, and the capacity to predict the environment \cite{Jacob1998} rose up. These processes were tightly linked to the arrival of autonomous agents \cite{Woese2002, SmithMorowitz2016}, setting up the stage for evolution of complexity through natural selection \cite{Darwin1859, Dennett1995, Woese2002}---an algorithmic process itself \cite{Dennett1995,JohnstonLoiuis2022} deeply related to information theory \cite{Fisher1930, Gould2002, Frank2009}.

    A crucial question here is whether this process can generate open-ended novelty \cite{McMullin2000, BedauRay2000, RuizMirazoMoreno2008, Day2011, TaylorMcMullin2016, deVladarSzathmary2017} and what kind of universal patterns might be involved \cite{CorominasSole2018}. In this context, a key question is what drives the rise of organismal complexity beyond simple replicators. One view sustains that selection has no intrinsic bias towards complex life forms \cite{Gould2011}. According to this view, bacteria constitute the peak performance dominating the evolutionary landscape. Organisms much simpler would not be viable, thus prevailing deviations from the peak fitness would, in average, look like an increase in complexity even if no bias favoring greater complexity exists. A naive rendering of this hypothesis suggests a single-peaked distribution of abundance across the spectrum of life's complexity. Is that the case? 
    
    Along with cognition \cite{LiquidBrains1}, parasites have been proposed as causal agents that boost biological complexity, grounded in a plethora of studies, particularly regarding viruses \cite{Iranzo2016,Koonin2017,SoleElena2018}. In general, parasites of very different nature are known to have a major impact on their host's survival \cite{Little2002}. This has inspired theoretical and computational efforts to explain how parasites can affect the genetic structure of host populations. In silico models \cite{Ray1991, LenskiAdami2003, Adami2006} have helped explore a series of essential questions concerning parasitic dynamics. These include the emergence of sex as a mechanism of resistance against parasites \cite{HamiltonTanese1990}, improving the efficiency of evolutionary algorithms \cite{Hillis1990}, or parasites as promoters of mutualism \cite{IkegamiKaneko1990}. {\em In silico} evolution models show how parasites can promote evolvability and the complexity of their hosts, an effect notoriously driven by the parasite's memory \cite{ZamanOfria2014}. These models share a common strategy of using digital genomes and simplified case studies to gain the highest insight. This logical description is an essential part of biology, and a computational picture of interaction can capture much more than we would initially expect \cite{Brenner1998, Nurse2008}. 
    
    In this paper we use such a formal description, with parasites and their hosts treated as abstract machines \cite{SeoaneSole2018}. Our model is inspired by the Turing machine model of computation \cite{Hopcroft1984} where an abstract computing device is used as a way to formalize computation in terms of a simple automaton that reads a binary tape. Despite its simplicity, this kind of formal approach has been used to explore the limits of information processing in both physical and biological systems \cite{Bennett1982}. Unlike that original framework, our ``Turing parasites'' deal with stochastic environments that they need to predict. Under the right conditions (which require sufficiently successful predictive power) they can replicate themselves. Moreover, they can contract or expand their computational capabilities if evolutionary change is allowed. As shown below, parasites can act as complexity enhancers in an open-ended manner. Our formulation of environments, hosts, and parasites as strings of bits with minimal rules brings our model very close to the language of computer science, information theory, and statistical physics. Our adoption of prediction as a driving principle shifts the focus to how a minimal (yet growingly complex) computation projects memory about the past into the potential future of an environment.

  \section{Methods}
    \label{sec:2} 

    \subsection{Minimal bit-guesser model}
      \label{sec:2.1}

      Bit-guessers $G$ are abstract machines that posses a model $\Gamma^G_E$ of an external environment $E$. An environment consists of a finite tape with ones and zeros, to which guessers access one bit at a time similarly to how Turing machines read their inputs---only, our guessers advance always in the same direction. Bit-guessers use their internal model $\Gamma^G_E$ (akin to a Turing machine's internal state) to attempt to predict the next environmental bits. Thus the tape is obviously read but, differently from Turing machines, in the current implementation our guessers do not {\em print} upon it---i.e.\ they do not modify their environments. In the following we introduce some notation about each of these elements. (See \cite{SeoaneSole2018} for a more thorough description. The Appendices contains an extended discussion of the model with a table that summarizes the most relevant symbols.) 
      
      \begin{figure*}[t]
        \begin{center}
          \includegraphics[width=16 cm]{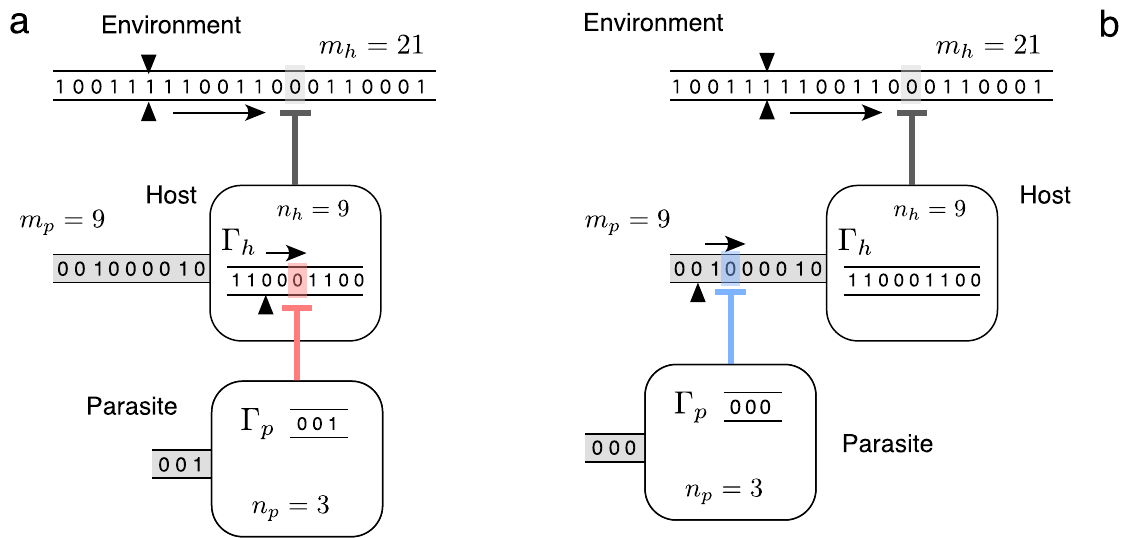}

          \caption{{\bf Host-parasite systems modeled as bit-guesser machines.} (a) A genotype parasite uses the host's internal model ($\Gamma_h$) as the environment off of which it extracts information to live. Note the reading head (light red) of the parasite interacting directly with the host's internal model about the external environment. (b) Phenotype parasites live out of anticipating the host's behavior---which is different from its internal model of the world. Note the reading head (light blue now) of the parasire now reacting to the external behavior that the host produces, while it does not interact with the host's internal model directly. Both host-based environments (its internal model and its behavior) have the same size, but phenotype parasites will usually encounter less complexity than genotype parasites. This is so due to the model's dynamics: A guesser's behavior is produced from its genotype by resetting each time that a mistake is made, thus making behavior more redundant than the model on which it is based. } 
      
          \label{fig:2}
        \end{center}
      \end{figure*}

      An $m$-environment ($E \equiv \{e_i; \> i=1, \dots, m\}$) consists of $m$ bits drawn randomly and uniformly. We evaluate an $n$-guesser in a given $m$-environment (with $n<m$) by dropping it at a random position, $i_0$, and asking that it predicts the next $n$ consecutive bits (Fig.\ \ref{fig:1}) with cyclic boundary conditions. These constitute an $n$-sized word $B(i_0) \subset E$. The guesser's internal generative model $\Gamma^G_E$ produces yet another $n$-sized word, $W^G_E(i_0)$, based on the guesser's memory and its past interaction with the environment (i.e.\ its history of correct and incorrect guesses as explained below). The fraction of bits correctly guessed reads:
        \begin{eqnarray}
          p^G_E(i_0) &=& {1\over n} \sum_{i=1}^n \delta(w_i, b_i). 
          \label{eq:1}
        \end{eqnarray}
      Here, we use Kronecker's delta $\delta(w_i, b_i) = 1$ if $w_i=b_i$ and $\delta(w_i, b_i)=0$ otherwise; and $w_i$ and $b_i$ are the $i$-th bit within $W^G_E(i_0)$ and $B(i_0)$ respectively. We might want to evaluate a guesser several times in a same environment (for which we drop it at different starting positions $i_0$) or in an ensemble of environments of fixed size $m$. For simplicity, we use $p^n_m$ to name the average number of bits guessed by an $n$-guesser in either $m$-environmental setup. 

      Bit-guessers pay a cost $c$ for each bit that they attempt to predict, and they rip off a reward $r$ for each correct prediction (Fig.\ \ref{fig:1}). Evaluating a guesser in an environment reports a net reward:
        \begin{eqnarray}
          \Delta \rho &=& (p^n_mr-c)n = (p^n_m - \alpha)rn, 
          \label{eq:3}
        \end{eqnarray}
      with $\alpha \equiv c/r$ measuring how meager an environment is: the larger $\alpha$, the less reward per correct guess. If $p^n_m < \alpha$, then $\Delta \rho < 0$ and the given $n$-guesser could not survive in that $m$-environment. In some of the numerical experiments that follow, if a guesser's accumulated reward $\rho$ is large enough, it gets replicated. 

      Our guessers were designed to have their computational complexity controlled by a single parameter---its size, $n$, which also imposes a replication cost. This computational complexity is implemented by the guesser's generative model, $\Gamma^G_E$, which consists of an $n$-sized bit-string used sequentially for predicting $E$. To elaborate $\Gamma^G_E$, we assume that the guesser has had access to the whole environment, and that it has come up with the best model possible constrained only by its size $n$ and a minimal correcting mechanism. The only information unknown to the guesser is where it will be dropped as it is evaluated later. Thus, $\Gamma^G_E(1)$ is just the most frequent bit in the environment, which is the most likely outcome if we drop it in a random position. After a correct guess of this first bit, moving onto the next position, the most likely guess $\Gamma^G_E(2)$ is the most frequent bit following every instance of $\Gamma^G_E(1)$ in the environment. We proceed building $\Gamma^G_E(i)$ as the most frequent bit following all ($i-1$)-sized words matching \{$\Gamma^G_E(1)$, \dots, $\Gamma^G_E(i-1)$\} in the environment. 

      As we evaluate a guesser, it proposes consecutively the bits in $\Gamma^G_E$ as long as its predictions are correct (Fig.\ \ref{fig:1}a). If there is a mistake (Fig.\ \ref{fig:1}b-c), the guesser resets to $\Gamma^G_E(1)$ for the next bit and proceeds onwards from there. More formally, the word $W^G_E(i_0; k)$ produced by guesser $G$ in the environment $E$ (Fig.\ \ref{fig:1}d) when it is dropped in $i_0$ is $W \equiv \{W(k), W(k) = \Gamma^G_E(k-l)\}$ where $k=1, \dots, n$ and $l$ is the last $k$ such that $W^G_E(i_0; k) \ne B(i_0-1+k)$. $l=0$ if no wrong guesses have occurred yet.

    \subsection{Genotype and phenotype parasites}
      \label{sec:2.2}

      Our model makes a distinction between the ideal best guess $\Gamma^G_E$ and the bits actually emitted as a guesser is evaluated $W^G_E$. We liken these to genotypes and phenotypes. $\Gamma^G_E$ stores instructions for prediction and dictates the agent's behavior expressed by $W^G_E$. Parasites making a living off of another organism can do so by predicting the host's inner structure or its behavior. In the real world, gene matching as well as external trait recognition can be used by parasites to recognise their targets. In order to consider these two scenarios, we study both and label as {\em genotype} and {\em phenotype parasites}, respectively. The former take the host's $\Gamma^G_E$ as their environment (Fig.\ \ref{fig:2}a), while the later dwell on the host's external features (such as behaviour) $W^G_E$ (Fig.\ \ref{fig:2}b). Because bit-guessers distill correlations from their environments, both $\Gamma^G_E$ and $W^G_E$ are more predictable than completely random environments of the same size. It is simpler to predict $\Gamma^G_E$ and $W^G_E$, but in turn they shall provide less reward. 

      We now need to differentiate between (i) the complexity of the external (or host's) environment ($m_h$) and the host's complexity ($n_h < m_h$); and (ii) the complexity of the parasite's environment, which is always the host's size ($m_p = n_h$), and the parasite's complexity ($n_p < m_p$). For clarity, we omit the host and parasite subindexes ($h$ and $p$), and just put a bar over all variables referring to parasites. Thus we name: $m \equiv m_h$, $n \equiv n_h$ (which equals $\bar{m} \equiv m_p$), and $\bar{n} \equiv n_p$. Evaluating parasites as usual, we have: 
        \begin{eqnarray} 
          \overline{\Delta \rho} &=& ( \bar{p}^{\bar{n}}_{\bar{m}} - \bar{\alpha} ) \bar{r} \bar{n}. 
          \label{eq:4} 
        \end{eqnarray} 
      Note $\bar{\alpha} \equiv \bar{c}/\bar{r}$ controlling the reward per correct prediction for parasites, with $\bar{\alpha} \ne \alpha$ in general. This parasite reward is subtracted from the host, so equation \ref{eq:3} is modified into: 
        \begin{eqnarray} 
          \Delta \rho &=& \left( p^n_m - \alpha \right) rn - \bar{p}^{\bar{n}}_{\bar{m}} \bar{r} \bar{n}, 
          \label{eq:5} 
        \end{eqnarray} 
      We assume that all the reward taken away from the host is efficiently transferred to the parasite. Variations allowing leaks would not affect our results substantially.

  \section{Results}
    \label{sec:3}

    \subsection{Bit-guesser size versus other measurements of complexity}
      \label{sec:3.0}
      
      \begin{figure}[h]
        \begin{center}
          \includegraphics[width=0.85\columnwidth]{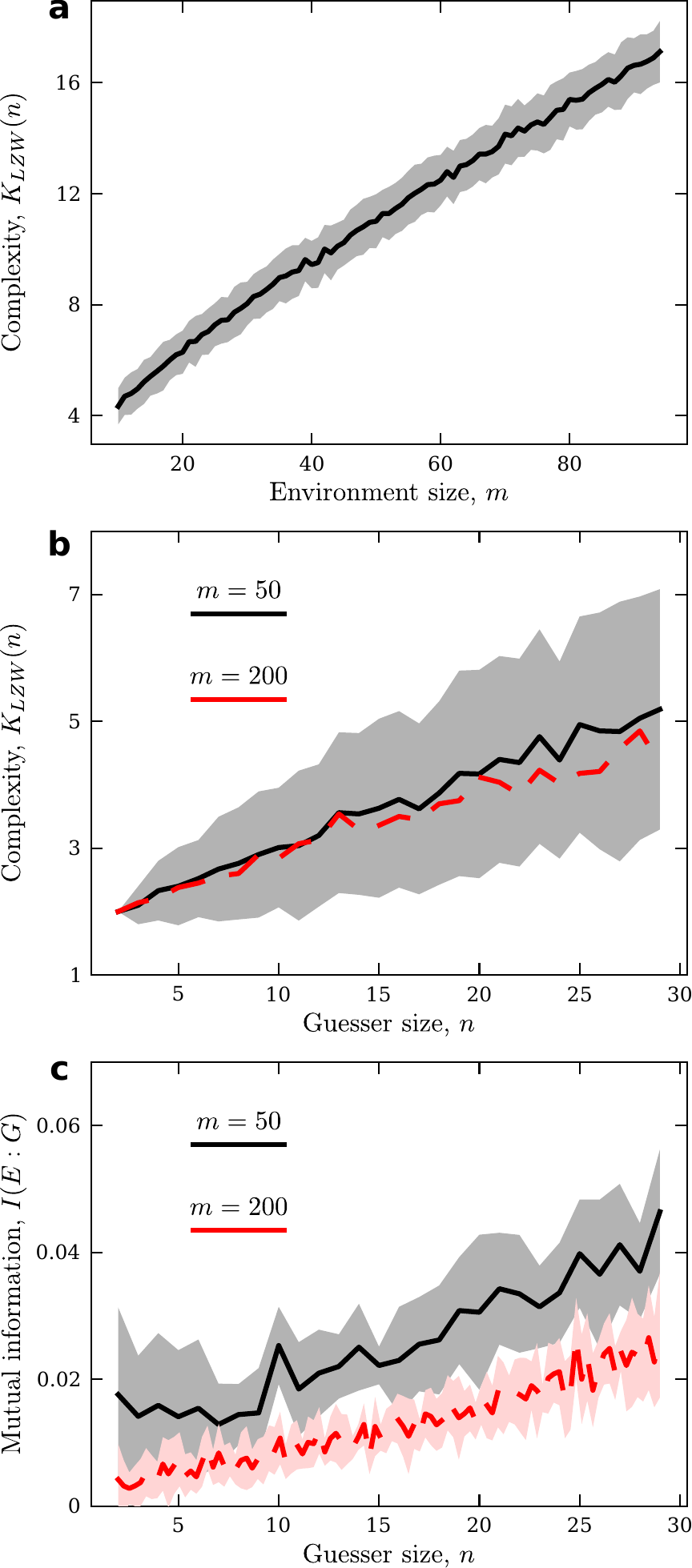}
  
          \caption{{\bf Comparing measurements of complexity.} (a) Environment complexity as a function of its size, $m$. Average over $100$ environments of each size, shading is standard deviation. (b) Same for guesser output as a function of its size, $m$. $1$ evaluation in each environment with $m=50$ (black) and $200$ (red, dashed; shading omitted). (c) Mutual information between environments and guesses. $10$ repeats of $1$ evaluation of the guesser in each of $100$ environments. }
  
          \label{fig:2.5}
        \end{center}
      \end{figure}

      Complexity is a thorny concept that eludes a clear-cut definition---we can find as many as authors \cite{Adami2002}. Many see this as a drawback for complexity science; but a similar caveat can be found regarding the definition of `life' and `living', and this has not prevented the advancement of biology---including in borderline cases such as the study of viruses. Besides having sensible measurements of complexity, we need to know how they relate to each other, understand what they are telling us mathematically, and assume that a same work can be seen from different angles. 

      Our model takes the sizes of environments and guessers as a measurement of their complexity. In the case of environments, this seems straightforwardly correct because they consists of periodically repeating strings of random bits. The longer an environment, the more variety and diversity we may encounter. In the case of guessers, we argue that their size is a good measurement of complexity because each additional bit confers them an expanded ability to find patterns within the environment. We made a deliberate choice to place all the guesser's algorithmic capability in its memory, specifically to have a single number comparable across guessers controlling complexity. Alternative implementations of guessers will result in different ways of quantifying procedural complexity, but similar results should follow as long as appropriate costs are placed on the expanded computational capabilities. One such example would be a guesser that, instead of one pattern (as in ours), stores a decision tree that is navigated whenever mistakes are made. In such a case, a measure of complexity should include the number of bits (like ours) but also the different branches and their depth, as well as placing costs on each decision. 

      Information theory offers ways to formalize these concepts, but not a straightforward manner of assigning costs to the different ingredients. Kolmogorov complexity \cite{Kolmogorov1965} is the length of the shortest program that can produce a given binary string. This measure is maximal for truly random strings, as the impossibility to predict the next bit from the previous history implies that any program producing the string must contain it in full. Completely regular strings (say, one that repeats `01' ad infinitum) have minimal complexity. In general, the Kolmogorov complexity of a string cannot be computed. Different correlates are used to approximate it in practical situations---for example, some notions of entropy calculated over the probability of finding different patterns within a string \cite{CoverThomas1999, CorominasSole2018}. Compression methods attempt to find minimal representations of a given string, thus they often approximate Kolmogorov complexity as well. The lossless Lempel-Ziv-Welch (LZW) algorithm \cite{ZivLempel1977,Welch1984} explicitly finds repeated patterns of variable length within a longer string to remove unnecessary redundancies. 

      Fig.\ \ref{fig:2.5}(a) shows the Kolmogorov complexity (as approximated by the LZW algorithm) for environments of increasing size. As expected, this grows monotonically with size---as $m$ grows, we encounter longer strings of random bits which require more and longer unique patterns to reconstruct. Thus environment size, $m$, is a good correlate of its Kolmogorov complexity. Fig.\ \ref{fig:2.5}(b) shows that guesser size, $n$, is also a good correlate of its corresponding complexity. This complexity seems to grow linearly with $n$, but it displays a much larger variance than environments. This suggests that the relationship between a Guesser's size and its Kolmogorov complexity can be highly non-linear and context dependent. 

      The crucial ability of a guesser to thrive comes from its capacity to correlate its bits with those of the environment, a property that we can capture with information theoretical and statistical measurements such as mutual information. In \cite{SeoaneSole2018}, we showed how this decreases, for guessers of a fixed size, as environments grow larger; and how it grows, for fixed environment sizes, as a guesser's complexity increases. Fig.\ \ref{fig:2.5}(c) shows an updated view of this relevant information-theoretical measure.

    \subsection{Performance in environments of different sizes}
      \label{sec:3.1}
      
      \begin{figure}[]
        \begin{center}
          \includegraphics[width=0.45\textwidth]{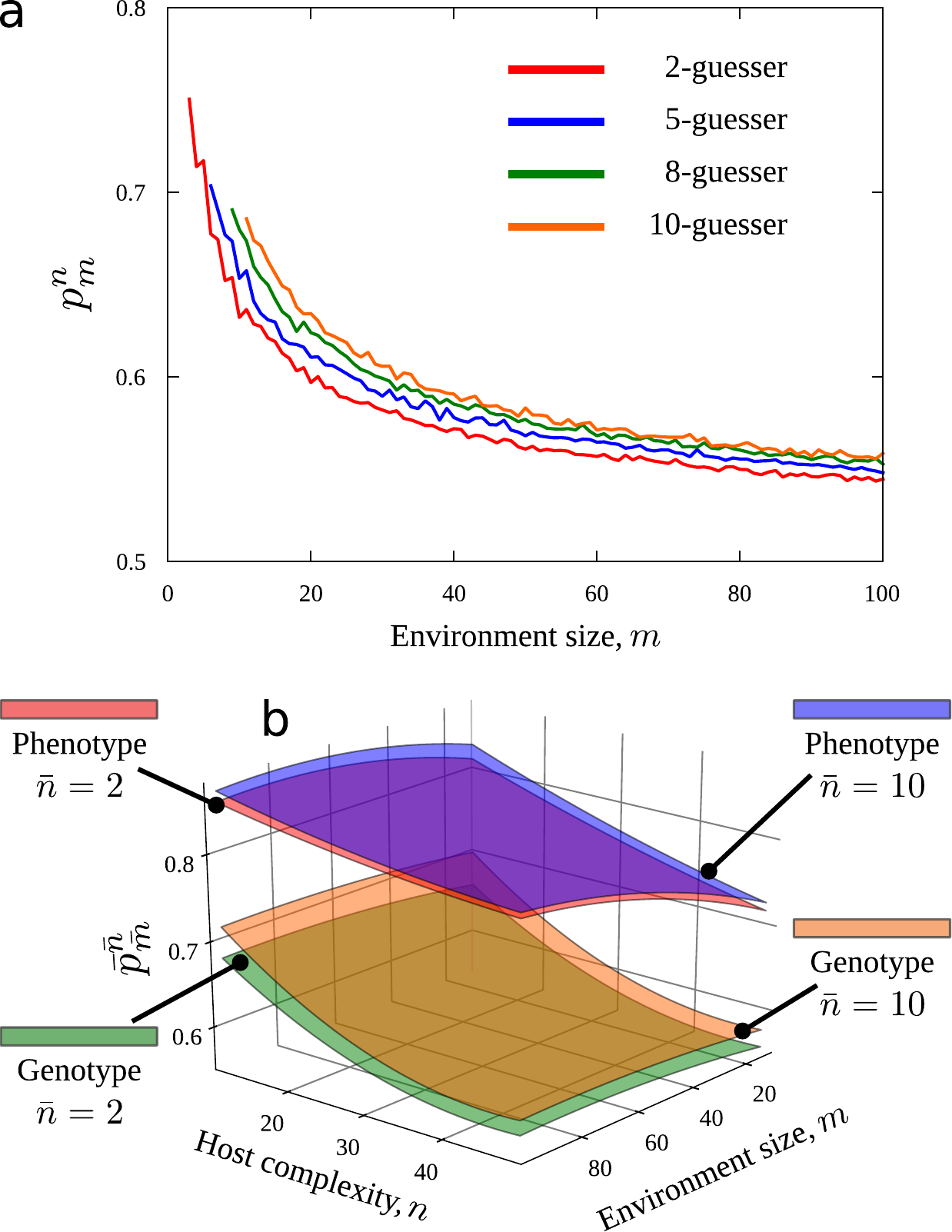}
  
          \caption{{\bf Host and parasite performances.} (a) Host's performance ($p^n_m$) as a function of its complexity and the environment's size. Curves are for 2-, 5-, 8-, and 10-guessers, with curves for bigger guessers laying on top and shifted to the right with respect to those for smaller guessers. (b) Parasite's performance ($\bar{p}^{\bar{n}}_{\bar{m}}$) as a function of both host's complexity and external environment's size. Shown are least square fits of quadratic surfaces to data from numerical experiments. }
  
          \label{fig:3}
        \end{center}
      \end{figure}

      Our simplest question is how many bits can hosts and parasites guess under specific circumstances. Expanding results from \cite{SeoaneSole2018}, Fig.\ \ref{fig:3}a shows $p^n_m$ for hosts. These curves delimit average survival conditions: $n$-guessers can survive in $(m, \alpha)$ combinations under their corresponding curve. More complex guessers (larger $n$) can survive in a wider range of conditions. For parasites, we vary both external environment ($m$) and host complexity ($n \equiv \bar{m}$). The volumes subtended by the surfaces in Fig.\ \ref{fig:3}b indicate combinations $(m, n=\bar{m}, \bar{\alpha})$ under which $\bar{n}$-parasites can survive. This volume is again bigger for more complex (larger $\bar{n}$) parasites. Survival regions decrease as host complexity grows (Sup.\ Fig.\ 1a in \cite{SupMat}), hence becoming more complex is a good strategy to diminish the survival chances of parasites. 

      Survival surfaces associated to phenotype parasites subtend much larger volumes than those of genotype parasites in Fig.\ \ref{fig:3}. Due to the reset mechanism in generating $W^G_E$, its complexity is bounded by that of $\Gamma^G_E$, thus making $W^G_E$ more predictable. Here our model differs from real organisms, in which genotypes are seeds that generate much more complex behavior. In any case, both $W^G_E$ and $\Gamma^G_E$ are easier to predict than external environments of the same size, which is the crucial advantage of parasitism.

    \subsection{Escaping parasites by increasing behavioral complexity}
      \label{sec:3.2}
      
      \begin{figure}[]
        \begin{center}
          \includegraphics[width=8.5 cm]{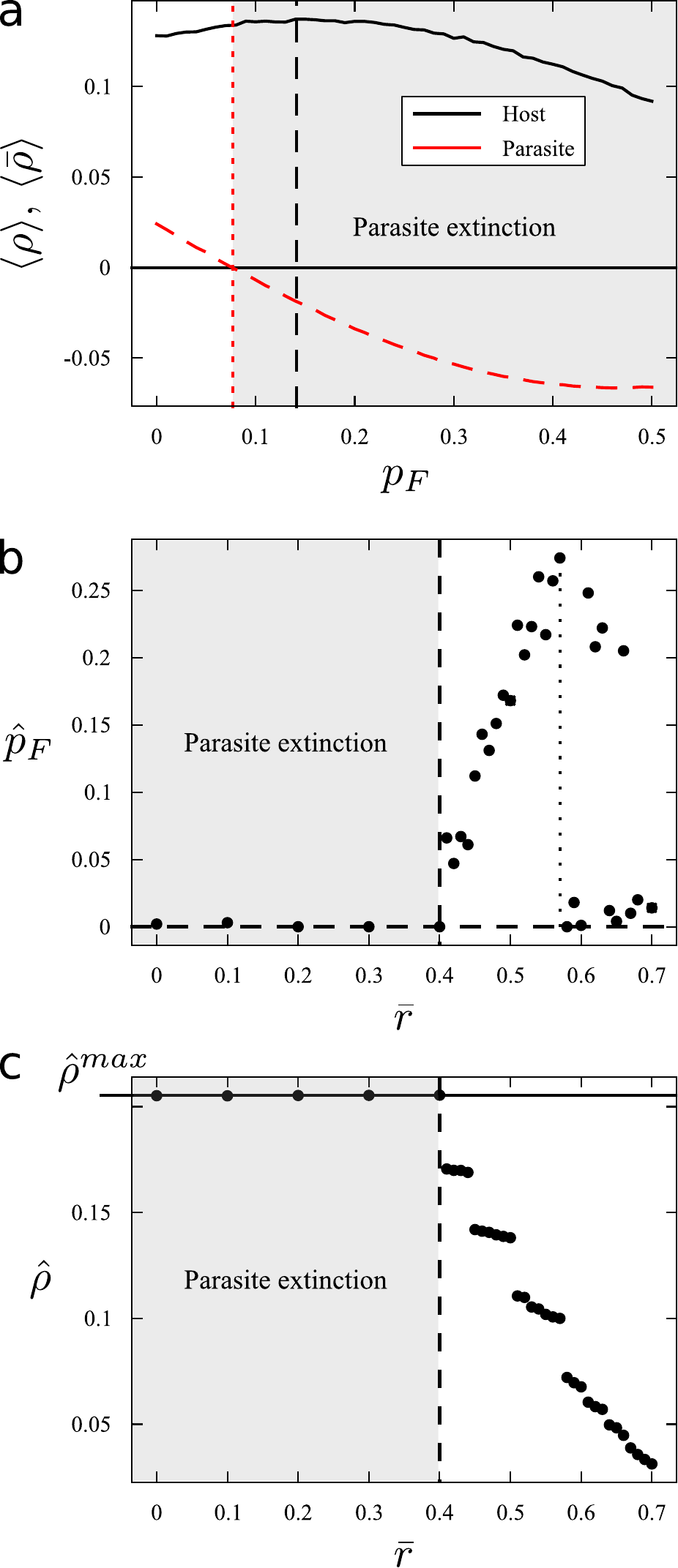}
  
          \caption{{\bf Escaping parasites by increasing behavioral complexity.} (a) Host ($n=10$, black solid curve) and parasite ($\bar{n}=4$, red dashed curve) average reward for $p_F \in [0, 0.5]$ ($m=30$, $\bar{r}=0.5$). For each $\bar{r}$, we evaluate $10^5$ hosts and parasites. If the parasite survives, $\bar{p}^{\bar{n}}_{\bar{m}} \bar{r} \bar{n}$ is subtracted from its host's reward. Optimal flip probability $\hat{p}_F$ (b) and reward $\hat{\rho}$ (c) reveal three different regimes: i) Low $\bar{r}$ (shadowed area), parasites die naturally, no additional behavioral complexity is needed ($\hat{p}_F=0$, $\hat{\rho} = \hat{\rho}^{max}$). ii) Intermediate $\bar{r}$, some additional behavioral complexity eliminates parasites ($\hat{p}_F>0$, which has a toll so that $\hat{\rho} < \hat{\rho}^{max}$). iii) Large $\bar{r}$, parasites endure, additional complexity does not help, ($\hat{p}_F=0$ again, $\hat{\rho} \ll \hat{\rho}^{max}$ because of the lasting parasite). }
  
          \label{fig:4}
        \end{center}
      \end{figure}

      A straightforward way to become more unpredictable is by adding some randomness (hence computational complexity, as rigorously defined \cite{Solomonoff1960, Chaitin1977, Chaitin1990, Kolmogorov1965}) to a planned behavior. We implemented hosts with their usual $\Gamma^G_E$, but who flip each bit in $W^G_E$ with probability $p_F$. Fig.\ \ref{fig:4}a shows how the net reward changes for hosts ($n=10$) and parasites ($\bar{n}=4$) as $p_F$ varies. The effect depends on the environment size (here $m=30$) and the parasite's share of reward ($\bar{r} = 0.5$). In Fig.\ \ref{fig:4}a, when $p_F$ is low, parasites survive, resulting in less reward for the host. For large enough $p_F$, the parasite is unable to predict its host and dies. But the randomness introduced entails that the host fails to predict some of the bits that it could, thus very large $p_F$ has a toll eventually. A tradeoff emerges between the parasite and environmental pressures resulting in an optimal level of randomness ($\hat{p}_F$, Fig.\ \ref{fig:4}b) for the host. The host's reward at this optimal value ($\hat{\rho}$, Fig.\ \ref{fig:4}c) can be diminished by either factor. 

      For $0 < p_F < \hat{p}_F$ the strategy partly thwarts the parasite's development. This would result in a slowed-down epidemiological spread. At the level modeled here, arrested (yet surviving) parasites have a smooth, parsimonious effect on the host. This regime might extend to $p_F > \hat{p}_F$ values, as $\hat{p}_F$ marks the threshold at which further random behavior by the host does not pay off, which is compatible with residually surviving parasites. 

      If parasites take away very little reward (e.g.\ $\bar{r}=0.4$, Sup.\ Fig.\ 3a in \cite{SupMat}), they cannot survive even in normally behaving hosts; thus $\hat{p}_F = 0$. This regime persists for a range $\bar{r} < \bar{r}^*$ (Fig.\ \ref{fig:4}b, shaded area). If $\bar{r}$ is very large (e.g.\ $\bar{r} = 0.7$, Sup.\ Fig.\ 3d in \cite{SupMat}), parasites survive even in fully unpredictable hosts. Thus flipping bits does not help the host against the parasite and still results in worst prediction of the external environment---hence $\hat{p}_F = 0$ again. Only $\hat{p}_F > 0$ for an intermediate range of $\bar{r}$ (Sup.\ Fig.\ 3b-c in \cite{SupMat}). Even if parasites are driven to extinction in this regime, the host misses some of the potential reward due to its degraded ability to predict (Fig.\ \ref{fig:4}c).

    \subsection{Increase of complexity in neutral ecosystem dynamics}
      \label{sec:3.3}

      We model neutral ecological interactions (Fig.\ \ref{supFig:1}a-c) with ecosystems that contain a number of spots. Each spot can be occupied by an $n$-guesser (restricted to $n = 1, \dots, 10$ in the next example) or empty (a $0$-guesser). An ecosystem presents fixed environment size ($m$) and harshness ($\alpha$). All guessers are initially endowed with a reward $\rho(t=0) = n\rho_0$, representing a satisfied metabolic load that grows with the guesser's complexity.

      To simulate ecosystem evolution, at every time step a spot is randomly chosen. The corresponding guesser is evaluated on a newly generated $m$-environment (Fig.\ \ref{supFig:1}a). As before, evaluation starts after the guesser has elaborated the best internal model ($\Gamma$) given its capabilities ($n$); then it is dropped on a random position of the environment and proceeds as in Fig.\ \ref{fig:1}. The balance from this evaluation is added to the guesser's accrued reward: 
        \begin{eqnarray}
          \rho(t + \Delta t) &=& \rho(t) + (p^n_m - \alpha)rn. 
          \label{eq:6}
        \end{eqnarray}
      If $\rho(t + \Delta t) > 2n\rho_0$, the guesser gets replicated (Fig.\ \ref{supFig:1}b) and an amount $n\rho_0$ (as initial endowment for the daughter) is subtracted: 
        \begin{eqnarray}
          \rho(t + \Delta t + \delta t) &=& \rho(t + \Delta t) - n\rho_0. 
          \label{eq:7}
        \end{eqnarray}
      The guesser keeps replicating until $\rho(t + \Delta t + \delta t) < 2n\rho_0$. Daughters are allocated to random spots in the ecosystem, which might be empty or not---in which case the older guesser is replaced. If, after being evaluated, $\rho(t + \Delta t)<0$, the selected guesser dies and a $0$-guesser occupies its spot (Fig.\ \ref{supFig:1}c).

      \begin{figure}[]
        \begin{center}
          \includegraphics[width=\columnwidth]{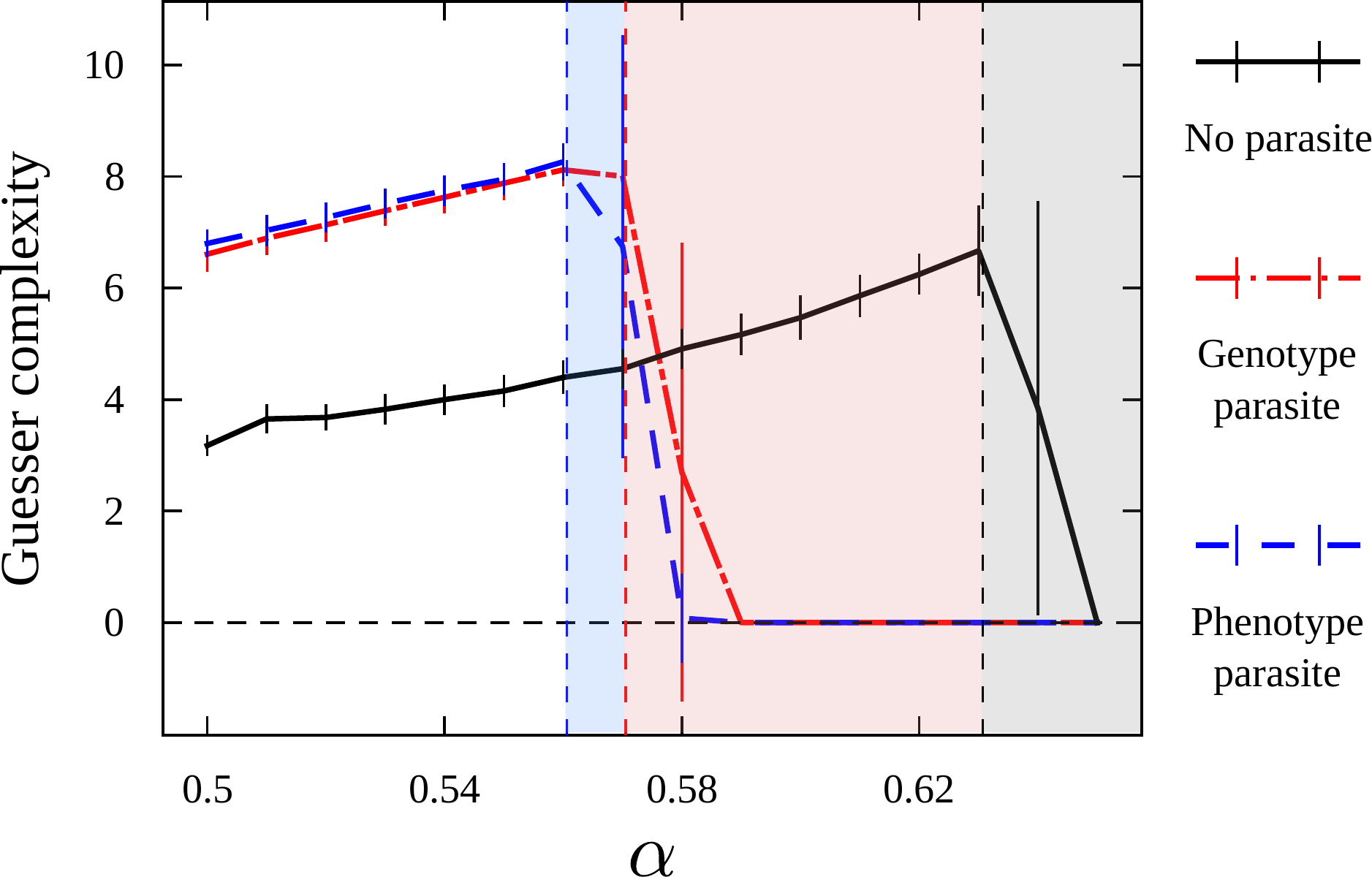}

          \caption{{\bf Ecosystem dynamics.} Average complexity of surviving guessers in an ecosystem after $200$ generations for a range of $\alpha \in [0.5, 0.65]$ for experiments without a parasite (black, solid line) and for ecosystems with genotype (red, dot-dashed) and phenotype (blue, dashed) parasites. Shaded areas mark ecosystem collapse as environments become too challenging for all guessers present. Parasites always have $\bar{n}=1$ in this example. The effect of such simple parasite is huge. Note, however, that we are exploring precisely the region of parameter $\alpha$ that separates persistence from extinction for the studied guessers (as seen in Fig.\ \ref{fig:3}a for $m=30$). }

          \label{fig:5}
        \end{center} 
      \end{figure}

      We evolved ecosystems with $1000$ spots, initially populated by random, uniformly distributed guessers with $n \in [1, \dots, 10]$. Fig.\ \ref{fig:5} (black, solid line) shows the average guesser complexity after $200$ generations (each generation consists in the sequential evaluation of $1000$ spots chosen at random, so in average every spot is evaluated once per generation) for $m=30$ and $\alpha \in [0.5, 0.65]$. In copious environments (low $\alpha$, large reward per correct guess) a few right predictions report large benefits. Simpler guessers replicate faster because of their lower replication threshold (proportional to $n\rho_0$). If a cheap, sloppy strategy provides enough reward, investing in costly computations is unfavored. But in austere environments (larger $\alpha$) simple guessers cannot accrue reward fast enough and a complex computational machinery pays off. Very meager environments ($\alpha \rightarrow 1$) become prohibitive for all guessers---as indicated by the sudden fall of complexity in Fig.\ \ref{fig:5}. 

      To each guesser, we now add a parasite that does not die nor replicate. It just sucks off reward according to equation \ref{eq:5} whenever its host is evaluated. Even the simplest ($\bar{n}=1$) parasite, either at the genotype (Fig.\ \ref{fig:5}, red dash-dotted line) or phenotype (blue dashed line), suffices to achieve a remarkable complexity boost. Simpler hosts are more predictable, hence parasites extract more reward---favoring complex hosts. In return, the point at which the ecosystem collapses happens for lower $\alpha$. Parasites are double-edged swords that drive complexity but can precipitate their host's extinction. This collapse happens slightly earlier for phenotype parasites---as expected, since, as discussed above, the complexity (and hence predictability) of $W^G_E$ is bounded by that of $\Gamma^G_E$.

    \subsection{Open-ended Red Queen dynamics}
      \label{sec:3.4}

      To study parasitic pressures in an eco-evolutionary setup we add a shadow ecosystem where parasites of different complexity dwell and undergo replication and death dynamics (Fig.\ \ref{supFig:2}). Replicating hosts and parasites can now produce simpler or more complex daughters through mutations. 

      All guessers replicate as explained above if 
        \begin{eqnarray}
          2n\rho_0 \le \rho(t+\Delta t) < (2n+1)\rho_0
        \end{eqnarray}
      Moreover, if the inequality 
        \begin{eqnarray}
          \rho(t+\Delta t) \ge (2n+1)\rho_0
        \end{eqnarray}
      holds, a mutation happens with probability $p_\mu=0.5$. (This relatively high mutation rate was chosen to illustrate model dynamics within a reasonable time and without using too many computational resources. Similar results as follow are obtained with samller mutation rates, but the interesting dynamics that we report below take longer---thus consuming much more computing time.) This mutation goes in either direction ($n \rightarrow n+1$ or $n \rightarrow n-1$) with equal chance. The corresponding initial endowment for the daughter is subtracted from the mother. When a host gets replicated, its parasite (if any) gets replicated alongside and occupies the corresponding spot in the shadow ecosystem (replacing an older parasite if necessary). The initial endowment of this daughter parasite is subtracted from the daughter host. Note that parasites also replicate through the usual route---i.e.\ their reward overcoming a threshold 
        \begin{eqnarray}
          \bar{\rho}(t+\Delta t) > 2\bar{n}\bar{\rho}_0. 
        \end{eqnarray}
      When this happens, the daughter parasite substitutes the guesser in a random spot of the shadow ecosystem. Parasites might also die, leaving a spot of the shadow ecosystem unoccupied. Thus, a host's parasite might change over time or disappear. The current model assumes $\bar{n} < \bar{m} \equiv n$, thus if a daughter parasite becomes too big (e.g.\ because it is allocated to a spot with a small host or with no host at all), she is promoted to host into the main ecosystem. 

      \begin{figure*}[]
        \begin{center}
          \includegraphics[width=\textwidth]{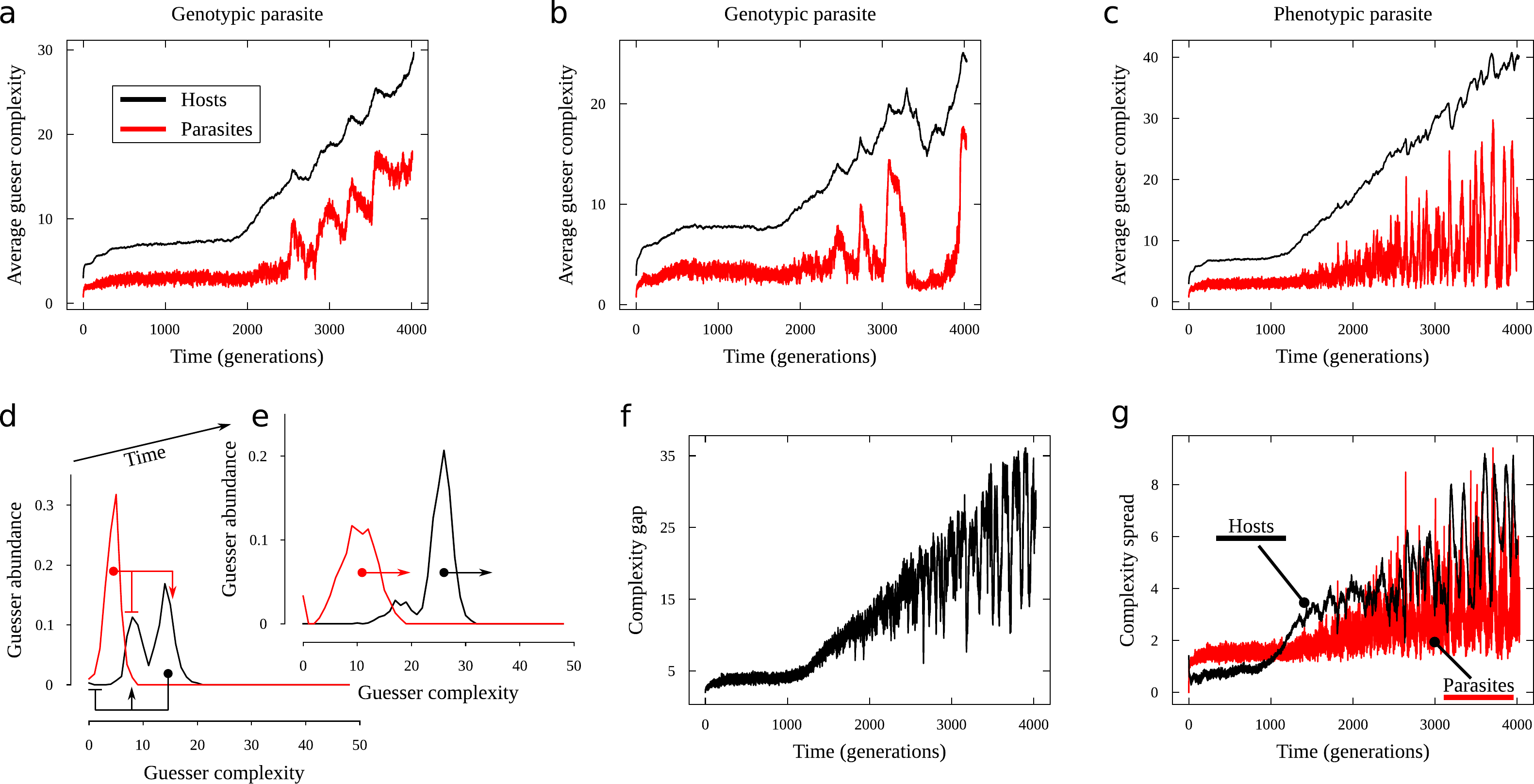}

          \caption{{\bf Red Queen dynamics in host-parasite co-evolution.} (a-c) Single run of a time evolution of average complexity of hosts (black, top curve) and parasites (red, bottom curve) (ecosystems have $1000$ spots, $m=50$, $\alpha = 0.5$, $r=1$, $\rho_0=1$, $\bar{\alpha}=0.6$, $\bar{r}=0.35$, $\bar{\rho}_0=0.005$, and $p_{\mu}=0.5$). Genotype parasites (a-b) present a more varied temporal unfolding than phenotype parasites (c). (d-e) Abundance of guessers of different complexity between hosts (black, right-most distributions) and parasites (red, left-most distributions) at two different times. Simpler hosts are easier to predict by their parasites, more complex hosts get away. Simpler parasites fail to cope with their hosts, while more complex parasites fare better. Effectively, simpler guessers are repressed and more complex ones are favored. (e) This results in a pair of effective forces pushing both host and parasite communities towards ever-higher complexity. (f) Average complexity gap between hosts and parasites of the simulation in (c). (g) Standard deviation of complexity of host (black) and parasite (red) populations at each given time of the simulation in (c)---not to confuse with the wide variation of the average population complexity over time. }

          \label{fig:6}
        \end{center}
      \end{figure*}

      Despite the simplicity of the model, it resulted in an unexpected range of behaviors---of which we only show the most salient ones. For certain parameters, the host-parasite eco-evolutionary interactions ignite Red Queen dynamics \cite{VanValenRic} in which both counterparts engage in a race of growing complexity, potentially without an end point, suggesting open-ended evolution (Fig.\ \ref{fig:6}a-c). All simulations were chosen such that, in the absence of parasites, the host ecosystem would be fully populated by $1$-guessers. This is, the observed trajectories are truly outstanding. Furthermore, $100$ repeats of each experiment have been carried out for a broad range of model parameters, robustly finding that Red Queen dynamics ensue sooner or later for many parameter settings (Sup.\ Fig.\ 4 in \cite{SupMat} illustrates complexity evolution for the $100$ experiment repeats with different parameters settings). 

      Fig.\ \ref{fig:6}d illustrates the mechanism behind the emergence of these Red Queen dynamics. Simpler hosts are easy to predict, while more complex ones escape their parasites. On the other hand, more complex parasites promptly thrive, while simpler ones fail to accrue reward and replicate. This results in a pair of effective forces pushing hosts and parasites towards ever increasing complexity. We halted our simulations if $n$ became larger than $m$, but our experiments do not show signs of slowing down. There is no principled reason why this mechanism could not operate indefinitely.

      Our simulations show other noteworthy, unexpected phenomena. We observed long periods of stasis (Fig.\ \ref{fig:6}a-c), suggesting relatively stable attractors which, anyway, are eventually escaped. Genotype parasites (Fig.\ \ref{fig:6}a-b) present more diverse temporal trajectories. Some show uninterrupted, relatively monotonous complexity buildup (Fig.\ \ref{fig:6}a). Others show ecosystem-wide complexity collapses (Fig.\ \ref{fig:6}b) into an alternative, seemingly meta-stable attractor with complex hosts and simple parasites. These states can persist for hundreds of generations, but are eventually escaped. The time trajectories of phenotype parasites (Fig.\ \ref{fig:6}c) are less diverse. In Sup.\ Fig.\ 5 in \cite{SupMat} we show how a range of behaviors can be achieved by varying just one parameter. Some of these dynamics are outstanding---e.g.\ punctuated equilibrium for very low $\bar{\rho}_0$. These behaviors were never purposefully hand-wired into our minimalist model.

      The complexity gap between hosts and parasites grows linearly during Red Queen dynamics (Fig.\ \ref{fig:6}f). Perhaps host and parasite complexity must observe some algebraic relationship to enable open-ended evolution---e.g.\ if hosts get too complex, they might kill off all parasites and halt the open-ended dynamics. 

      Notice that fluctuations in complexity grow as all guessers become more complex. This happens both as fluctuations of the average population complexity (and most notably for phenotypic parasites, as Fig.\ \ref{fig:6}(g) illustrates), and as spread within the population at each time (Fig.\ \ref{fig:6}g). This panel shows the standard deviation of guesser sizes within the ecosystem at each time. We observe that this variance is larger for parasites during the stale phase, but that it becomes larger for hosts during the Red Queen dynamics. Both hosts and parasites present: (i) large fluctuations in the variance of their complexity and (ii) an overall trend of increasing standard deviation of complexity. The fluctuations indicate large, sudden contractions and expansions of the distribution of guesser sizes, suggesting non-trivial eco-evolutionary interactions. The overall increasing standard deviation of the complexity, on the other hand, indicates that not only are both hosts and parasites becoming more complex over time, in average; but also both populations themselves are becoming more heterogeneous over time. This suggests that, as the Red Queen dynamics proceed, more combinations of host-parasite couples might become viable. 

      Finally, we could think that more complex parasites might be lethal and result in emptied ecosystems (as it happens without evolutionary dynamics for large $\alpha$, Fig.\ \ref{fig:5}). Sup.\ Fig.\ 6 in \cite{SupMat} shows, instead, that ecosystem occupation remains at its maximum for hosts and increases lightly over time for parasites.

  \section{Discussion}
    \label{sec:4}

    It has been conjectured that parasites act as a pressure for more complex hosts. Here we sought the simplest mathematical description that captures this qualitative hypothesis. We built upon the bit-guesser model \cite{SeoaneSole2018} as a minimal framework in which complexity (as grounded in information theory and computer science) is parsimoniously connected to Darwinian selection by organisms that thrive and replicate if they successfully predict their environment. These elements (replication, selection, and information processing), among others, set biology apart from inert matter \cite{Hopfield1994, SzathmaryMaynardSmith1997, Joyce2002, WalkerDavies2013, SmithMorowitz2016}. 

    \begin{figure}[h]
      \begin{center}
        \includegraphics[width=7.5 cm]{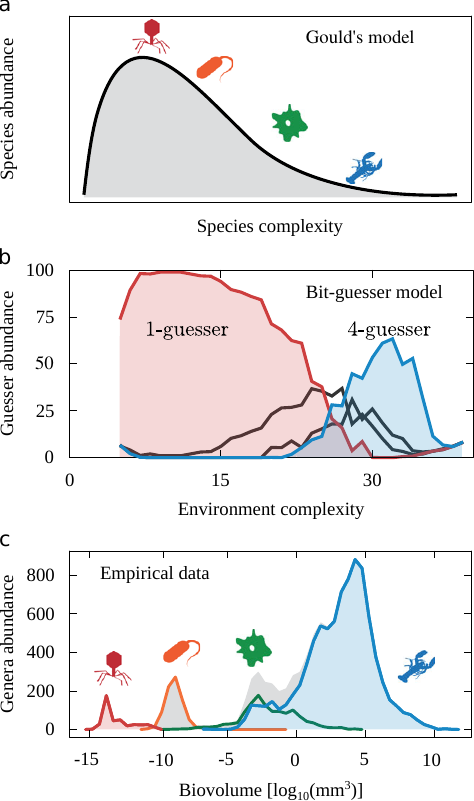}

        \caption{{\bf Different models of organismal complexity across scales.} (a) A proposal by Gould \cite{Gould2011} that more complex life arises through sheer random drift leads to a unimodal distribution of life complexity peaked around the {\em most successful} class of organisms (bacteria, according to \cite{Gould2011}). This is a cartoon distribution modified from \cite{Gould2011, HeimWang2017}. If evolutionary pressures for the emergence of complex life exist, it becomes more likely that steps of organismal complexity occupy their niches and present multi-peaked distributions across scales. Some such evolutionary drivers were numerically characterized in \cite{SeoaneSole2018} and produce multipeaked distributions ((b), where abundances show percentage of ecosystem occupation as in Sup.\ Fig.\ 6 in \cite{SupMat}). Multipeaked distributions are also derived from exhaustive data bases \cite{HeimWang2017} (c). }

        \label{fig:7}
      \end{center}
    \end{figure}

    Our minimalist model shows that: (i) increased behavioral complexity can be a valid strategy to scape parasites; (ii) more complex organisms result easily from the introduction of simple, immutable parasites in an ecosystem; and (iii) eco-evolutionary dynamics can result in Red Queen dynamics of hosts and parasites becoming more complex to scape each other. These results join other explicit drivers of increased biological complexity \cite{SeoaneSole2018}, as well as evidence from large census of species (Fig.\ \ref{fig:7}c) \cite{HeimWang2017}. Together, they weaken more neutralist views \cite{Gould2011} which propose that there are not explicit evolutionary pressures favoring higher complexity in biology. It is reasonable to think that these forces (notably parasitism) have been operating since very early in the history of life. This offers reassuring arguments that complex life is expected under an array of circumstances, increasing the likelihood that organisms such as higher metazoans or advanced cognitive systems did not arise by random drift.

    The emergence of Red Queen dynamics is our most important finding. A wide set of conditions lead to a quick, seemingly open-ended, surge of complexity. Our results imply that powerful forces underlying common biological interactions will drive life towards great complexity under the adequate conditions. In our opinion, this robust mechanism turns the original question (``are there pressures towards complex life in Darwinism'') on its head. We should now wonder what the consequences of this strong evolutionary dynamics can be in the real world, or under what circumstances this mechanism can be attenuated or harnessed. In this last regard, we say nothing about the explicit implementation through which this complexification can be achieved. 
    
    Our model is limited by finite size constraints, but the complexity increase did not show a tendency to stop or saturate. There is no reason why the principles operating at the studied scales should not work indefinitely, suggesting that host-parasite dynamics might drive open-ended evolution. This feature can be suggested from our model, that shows much longer and more sustained increases in complexity than other models \cite{ZamanOfria2014, HickinbothamHogeweg2021} and it occurs at seemingly constant rates. This suggests that, complexity does not affect much the rate of evolution. Even within the limited range studied, a very rich array of behaviors emerge. Noteworthy are the observed collapses of complexity (Fig.\ \ref{fig:6}b), which suggests that forward complexity evolution requires hosts and parasites progressing apace; and the punctuated equilibrium observed for very small parasite replication threshold ($\bar{\rho}_0$, Sup.\ Fig.\ 5e-f in \cite{SupMat}). These features were not anticipated. The minimalism of our model suggests that such phenomena might be general within host-parasite co-evolution. The fact that the collapses do not always happen (Fig.\ \ref{fig:6}a) and that they can be overcome (Fig.\ \ref{fig:6}b) shows that they are not an upper limit of complexity (which does not prevent it from existing anyway). 

    Multiple biological examples of host-parasite coevolution \cite{Sagan1967, BonenDoolittle1975, DawkinsKulski1999, MiKeith2000, VillarrealDeFilippis2000, Bell2001, HughesCoffin2001, Whitfield2002, Forterre2002, BlaiseHeidman2003, MartiLanzrein2003, WylerLanzrein2003, EspagneDrezen2004, KiepielaZimbwa2004, Claverie2006, ClaverieFournier2006, Bell2009, ForterrePrangishvili2009, Lane2010, Koonin2011, Witzany2012, SundaramWang2014, Lane2015, ChuongFeschotte2016, ChuongFeschotte2017, CornelisHeidmann2017, ChaikeeratisakVilla2017, VillarrealRyan2018, MizuuchiIchihashi2022} (notably those involving genomes) present elements very reminiscent of computer science and information theory (e.g. operations based on pattern matching, copying, etc.). Thus, despite the abstraction of our model and results, it is not unreasonable to try comparing them to empirical data. This paper is a first step in that direction. We derived quantitative bounds to computational aspects of host-parasite interactions, similar, e.g., to thermodynamic bounds for computation that apply in physical and biological systems \cite{Landauer1961}. Such bounds are better embodied by the $\bar{p}^{\bar{n}}_{\bar{m}}$ surfaces in Fig.\ \ref{fig:3}b. Further qualitative observations relevant for potential empirical studies are the gaps observed between host and parasite complexities. Might this gap also grow linearly in real host-parasite systems? To asses this we need to tackle complexity in real biological systems---a difficult issue \cite{Bonner1988}. Genome size is a first proxy, but examples of simple organisms with large genomes abound. The Kolmogorov complexity \cite{Kolmogorov1965} of a genome might remove redundancies and get us closer to the amount of useful information encoded. Beyond this, protein, functional, and behavioral repertoires might be our next approximations to capture the complexity of living beings. 

    Here we focused on a specific, biologically inspired view of parasitism. However, our mathematics demand only two populations of co-evolving agents with one of them making a living out of predicting the other, upon which a small damage is inflicted. The phenomenology that we uncover should be relevant for any situation fulfilling these conditions. Within biology, certain aspects of male-female interactions have been framed as host-parasitic relationships, with males of some species (notably among fish) openly described as parasitic. Our results would imply that the split in two sexes could, in certain cases, be yet another powerful engine for fast biological complexification. However, additional aspects should be taken into account (e.g.\ shared descent between hosts and parasites) which might modify the dynamics. 

    An important feat in the early evolution of life is the transition to a code capable of Darwinian evolution (and the computational processes that this demands) \cite{Eigen2000, Davies2019}. This acquires a greater importance in metabolism-first models of the emergence of life. Before an actual code exists, such models often rely on autocatalytic cycles as a self-replicating structure with Darwinian dynamics, without a centralized encoding of the information. What might have prompted the computational complexification of these structures in the first place? It turns out that autocatalytic cycles are affected by their own parasites as well \cite{AmorSole2017, BoerlijstHogeweg1991, SardanyesSole2007}. These are molecules or metabolic pathways that benefit from the catalytic activity without implementing any step of the cycle (i.e.\ they are metabolic drains). Under the light of our results, such models would be equipped with a ready-made evolutionary pressure for increasing computational complexity from the very beginning. The emergence of new parasites would be dictated by the exploration of new metabolic pathways by the autocatalytic set itself. Phenomena such as the periods of stasis that we observe would provide ratcheting platforms that would prevent mounting complexity from fading away. 

    The essence of our framework might capture non-standard parasites in social, technological, or economic systems (e.g.\ as businesses profit from each other's intellectual efforts, as hackers exploit the computational power of unaware remote servers \cite{BarabasiBrockman2001}, or as traders anticipate each other's moves \cite{LewisBaker2014}). Parasites might also have shaped the evolution of neural structures \cite{delGiudice2019}, thus extending our questions to machine learning and cognition in general. The recent, outstanding success of Generative Adversarial Networks \cite{GoodfellowBengio2014, RadfordChintala2015} relies on two systems (while of fixed complexity) establishing antagonistic dynamics similar to ours: a network gains fitness by fooling the other with artificial data, and the other becomes fitter by learning to discern fabricated examples. Our results offer a window to study the emergence of increasingly complex representations in digital ecosystems, as well as serious hypotheses about drivers of advanced cognition in the real world.

\vspace{0.2 cm}

  \section*{Acknowledgments}

    We want to thank members of the Complex Systems Lab and Dr Amor from the Gore lab at MIT's Physics of Living Systems for useful discussion. Seoane wishes to acknowledge the important insights contributed by all members of RCC. This work has been supported by the Bot{\'i}n-Foundation by Banco Santander through its Santander Universities Global Division, a MINECO grant FIS2015-67616-P (MINECO/FEDER, UE) fellowship, and AGAUR FI grant by the Universities and Research Secretariat of the Ministry of Business and Knowledge of the Generalitat de Catalunya and the European Social Fund and by the Santa Fe Institute. Seoane was funded by the Spanish Department for Science and Innovation (MICINN) through a Juan de la Cierva Fellowship (IJC2018-036694-I).

\appendix 

  \section{The bit-guesser model, step by step}
    \label{app:1} 

    Our current work builds upon the bit-guesser model introduced in \cite{SeoaneSole2018}. This model seeks a minimal representation of self-replicating agents that must predict their environment to survive. In searching for minimalism, some conceptual licences are taken and the model turns rather abstract. This section contains an informal, yet more detailed discussion of the model in the hope that it will facilitate the understanding of its different elements. 





    \subsection{The environment} 
      \label{app.1.1}

      Environments in the bit-guesser model consist of a finite tape of size $m$ containing zeros and ones. This is inspired by Turing's original model of computing machines. Within our model we are usually interested in knowing the performance of a bit-guesser in sets of environments that constitute ensembles in a same class of equivalence. For example, we often want to know the average performance of a guesser in the ensemble $E^m$ consisting of all environments with size $m$. This becomes unrealistic for large enough $m$, so we are content to sample a subset $\hat{E}^m \subset E^m$. Note that two $m$-environments are usually different: one might have all $0$s and another one a balanced mixture of $1$s and $0$s. Environments in such an ensemble are usually generated on the spot as needed, and stochastically; thus at least some average properties are equivalent, especially when we evaluate a guesser over many replicas. 

      Environments have periodic boundary conditions, such that if we move along them, their last bit loops back to the first one. Both this periodicity and the finiteness of our binary tapes are choices to study how much useful information can be extracted from them. Instead of $m$-environments, we could have modeled an infinite tape with a random distribution of bits; and have this distribution modulated by a parameter that would introduce correlations between positions on the tape. This would result in an unbalance of {\em words} of different sizes, which would make certain patterns more predictable than others. The finiteness of our $m$-environments has an effect similar to such correlations. The mechanism is at follows: 
        \begin{itemize}
          
          \item In a short environment, we expect to find more deviations from an ergodic sampling of all possible distributions of $1$s and $0$s than in a larger environment. As $m$ grows, the chance that we generate a random environment with an excess of, say, $0$s becomes much smaller. Thus, for larger enough $m$, the most likely is that both $0$s and $1$s are equally represented, thus knowing the likelihood of the most frequent bit will be each time less useful to predict any bit in the environment. 

          \item The same would happen if we would look at the distributions of $2$-words. This is: in short environments it is more likely that we find an unbalance of $00$, $01$, $10$, or $11$. The balance is recovered as $m$ grows. However, since the distribution of $2$-words needs more information to be fully specified, an ergodic sampling is reached for higher values of $m$ than for $1$-words. This implies that $2$-words remain useful over a wider range of $m$-environments. 

          \item The same happens for distributions of longer words. On top of that, computational mechanisms (such as our bit-guessers) that can exploit the information of words of a given length can usually exploit information of shorter words---thus the benefits accumulate. 

        \end{itemize}

      There are other ways in which we could have introduced how ``meaningful'' (and hence worth learning) patters of different sizes aire. The finite tapes chosen offer an elegant way that parameterizes environment complexity with just an intuitive number, $m$.

    \subsection{The guessers} 
      \label{app:1.2} 

      Similarly to environments, there are several ways in which we could have modeled self-replicating agents with computational capabilities. We could have opted for recurrent neural networks, epsilon machines, Bayesian or Boltzmann networks, etc. But these options go against our search for minimalism. In conceiving bit-guessers, we were hoping to capture just the essential elements and, hopefully, to summarize a guesser's computational costs and capacities with as many parameters as possible. 

      We converged to a model that also consists of a tape of bits, plus a minimal `if' to allow for the simplest error correction possible. The internal model of an $n$-guesser ($\Gamma$) consists of $n$ sorted bits, now without boundary conditions. We can conceive an evolutionary dynamics of $n$-guesser-candidates over a given, fixed $m$-environment: First, let us produce a lot of $n$-guessers, each with its own model ($\Gamma_i$) of the environment generated randomly. Then, let us evaluate each of these guesser-candidates (as specified below) in the given, fixed $m$-environment. Next, let us select the best-performing $\Gamma_i$ and, from them, produce new guesser-candidates through mutation and cross-over of their internal models. As we iterate these steps, we would expect our population to converge to the best guessers possible given their $n$ bits of computational capability. These evolutionary dynamics might be very interesting on their own, but in our research we cared more about the limits to performance that $m$-environments allow given their own complexity. Therefore, we generated our bit-guessers with arguably the best guess that they could come up bit, given their computational capacity. 

      Once an $n$-guesser has been initialized with the best guess possible, it is evaluated by being dropped onto a random position of its $m$-environment, then requesting that it predicts the bit in that and each of $n-1$ consecutive positions. Hence, the first best guess possible given the $m$-environment is the most frequent bit in it. This constitutes the first bit ($\Gamma_1$) in $\Gamma$. If the guesser is dropped in a position whose bit matches $\Gamma_1$, the best guess for the next position is the bit that most frequently appears after each instance of $\Gamma_1$ in the environment. If, from the position where the guesser was dropped, the second bit also matches its model ($\Gamma_2$ now), the best guess for the third position is the bit that most frequently appears after each instance of the $2$-word $\{\Gamma_1, \Gamma_2\}$. This goes on, thus these assorted collection of bits (up to the $n$-th iteration) constitutes the best guess with which $n$-guessers are equipped. 

      In evaluating a guesser, it produces a behavior ($W$) that consists of all the bits that it comes up as a guess after it is dropped in a random position of the environment $E$ and advances forward over it. Thus, the first bit in $W$ will invariably be the first bit in $\Gamma$ as well. If this guess is correct, the next bit in $W$ will be the next bit in $\Gamma$ as well, and so on. If one of the guesses is incorrect, the $n$-guesser has not got a heuristic to infer {\em where} it is located within the environment---it got lost. The best it can do is to go back to the starting point, and guess that the next bit will be the most frequent one in the environment---hence the next bit in $W$ is the first bit in $\Gamma$ again, from which the guesser proceeds as before as long as it keeps proposing correct guesses. The reset mechanism when a guess is mistaken is an `if' command---the only one allowed to bit-guessers. 

      Two important aspects are worth considering: First, instead of this hard reset, we could have modeled inference machines that navigate a tree. If we would do this, additional complexity is introduced, as we now need to parameterize the width and depth allowed in the tree. This counters our search for minimalism. It would also not be obvious how to weight the cost of successive `if' statements with respect to memory. Second, when looking at bit-guessers behaving (it helps to write down a few examples), it might occur that the guesser appears ``too dumb'' not to find an obvious pattern. For example, a $2$ guesser could never detect the repeating pattern in the string $110110110110110110$. This is, precisely, what it means that the computational capacity of this guesser is limited by $n=2$. 

    \subsection{Numerical experiments with bit guessers} 
      \label{app:1.3} 

      In \cite{SeoaneSole2018} and in this manuscript we describe a series of experiments that involve guessers coexisting or competing in a modeled ecosystem. Our ecosystems consist of a set of $N$ spots, each of which can be occupied, or not, by a guesser. In all our examples $N=1000$. Ecosystems evolve in {\em generations}, with one generation consisting on the evaluation (and implementation of death and replication dynamics if necessary) of the guessers in $N$ spots. In all our experiments, ecosystems are identified by environments of a fixed value, $m$.

      \begin{figure}[]
        \begin{center}
          \includegraphics[width=\columnwidth]{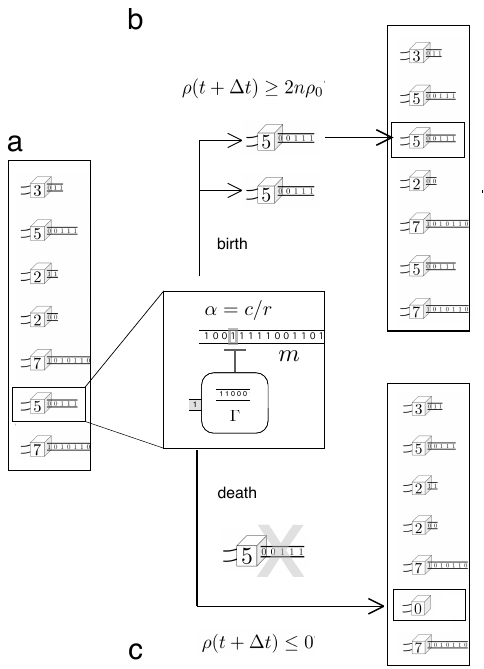}

          \caption{{\bf Ecological dynamics in a simple setup}. (a) From ecosystems with a finite number of slots, guessers are randomly picked up for evaluation against environments of a fixed size. (b) Good performance (i.e.\ good prediction of the environments) can get a guesser replicated. (c) Failing to anticipate the environment can get a guesser killed. }

          \label{supFig:1}
        \end{center}
      \end{figure}

      The simplest experiment with ecosystems is described in Fig.\ \ref{supFig:1}, and corresponds to experiments implemented in \cite{SeoaneSole2018}. In this experiment, the ecosystem is initialized by filling in the spots with random guessers that have $n \in [1, \dots, n^{max}]$, were $n^{max}$ delimits a range of explored guesser complexity\footnote{Here, $n^{max}$ marks a range of guesser complexity. Guessers with higher complexity are not allowed by design. This was used in \cite{SeoaneSole2018}, but it becomes irrelevant in the experiments of the current manuscript, as we will see.}. Once initialized, generations start running. A spot is randomly chosen for evaluation of the guesser that it contains. The likelihood of picking up a $1$ guesser is twice the likelihood of picking up a $2$-guesser, three times the likelihood of picking up a $3$-guesser, and so on. In this way, in average, guessers are offered a chance to guess a same number of bits. 

      The guesser picked up for evaluation (a $5$ guesser in Fig.\ \ref{supFig:1}) is presented with a newly generated $m$-environment. This environment is created randomly and on the spot, such that the ecosystem as a whole is evaluated over a subset of the $m$-environment ensemble ($\hat{E}^m \subset E^m$, as described above). The selected guesser first comes up with the best guess given the specific $m$-environment that has been generated. This implies two important things: i) We are again looking at limitations set up by the guesser complexity, as we decide to evaluate the best inference possible given the size, $n$, of the guesser (which determines its computational capabilities). ii) All $n$-guessers are equivalent to each other given the environment. This is so because we wish to find differences, hopefully, that are only due to the computational specifications of guessers and environments. Note, however, that two different $n$-guessers with a same $n$ evaluated at different times will be confronted with different $m$-environments---thus their best guesses would differ (because they are determined by the environment). Note also that if we would evaluate a same guesser twice on a same specific $m$-environment, while it would end up with a same internal model $\Gamma$, this might generate two different behaviors ($W$) because the later also depend on the position of the environment onto which the guesser is dropped. Note that $2$-guessers in Fig.\ \ref{supFig:1}a show different $W$.

      Guessers accumulate reward over time ($\rho(t)$) as described in the main text. If this reward is kept positive, the guesser survives. If this reward doubles a ``metabolic load'' ($\rho > 2n\rho_0$), the guesser gets replicated (Fig.\ \ref{supFig:1}b) and produces a new guesser with the same $n$. This new guesser occupies a random position of the ecosystem. This random position might be empty or occupied by another guesser. In this case, the older guesser is substituted by the new one. If $\rho(t) < 0$, the guesser dies (Fig.\ \ref{supFig:1}c). Its spot in the ecosystem is occupied by a $0$-guesser (i.e.\ an empty space). When computing the average complexity of an environment (i.e.\ the average $n$ of the guessers contained), empty spaces are scored as a $0$-guesser. This is how ecosystem complexity in Fig.\ \ref{fig:5} of the main text drops to $0$. 

      Besides $m$, experiments with simple ecosystems have fixed values of other model parameters such as $\alpha$ and $\rho_0$. Together, they determine what guesser complexity survives and which dies off. For example, very stringent environments have large $\alpha$ and they offer very little reward per guessed bit (as compared to the cost of attempting a guess). In these cases, more complex guessers (which make more correct guesses per attempt) might be favored. Furthermore, in the current paper we have introduced a static parasite to such environments (subsection III.C of the main text). This also affects what guessers survive, as shown in Fig.\ \ref{fig:5}. \\ 

      \begin{figure}[]
        \begin{center}
          \includegraphics[width= \columnwidth ]{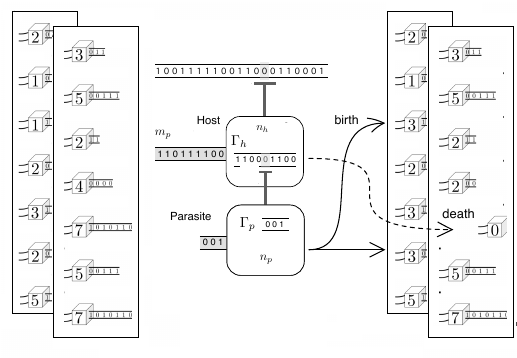}

          \caption{{\bf Eco-evolutionary dynamics}. Host-parasite dynamics in ecosystems are implemented through a shadow ecosystem such that parasites are picked for evaluation along with their hosts. }

          \label{supFig:2}
        \end{center}
      \end{figure} 

      More interesting experiments are introduced by the co-evolution of hosts and parasites---and they bring about the most alluring results in the paper concerning Red Queen dynamics. To implement these experiments we introduce what we called a {\em shadow} ecosystem, which is nothing but the possibility that each guesser in a main ecosystem (which we refer as hosts) might be infected each by a single parasite. Shadow ecosystems are depicted as a box behind the main ecosystem in Fig.\ \ref{supFig:2}. In this version of the experiments, host guessers are picked up for evaluation just as before (in Fig.\ \ref{supFig:2}, a $9$-guesser is selected). If there is a guesser in the same position of the shadow ecosystem, it is selected for evaluation as well. (In the same figure, there is a $3$-guesser parasite.) Evaluation of the host proceeds as usual. Evaluation of the guesser proceeds as explained in the Methods section of the main text, depending on whether we are dealing with a genotype or phenotype parasite. The parasites in Fig.\ \ref{supFig:2} affect genotypes. The reward acquired by the parasite is subtracted from the host's accumulated reward. 

      If the host's reward becomes negative ($\rho(t) < 0$), it dies and it is removed from the experiment, leaving an empty spot ($0$-guesser) behind. If the host was infected by some parasite (i.e.\ if there was a parasite in the same position of the shadow ecosystem), it dies as well (again, leaving a $0$-guesser behind in the shadow ecosystem). If the host doubles its replication threshold ($\rho(t) > 2n\rho_0$), it gets replicated. In doing so, it might displace an earlier host guesser as explained above. If the host carries a parasite, it gets replicated as well. The minimal reward needed to establish the daughter parasite ($\bar{n}\bar{\rho}_0$) is subtracted from the daughter host. It might seem that hosts are attached to their parasites forever, but a parasite might die because its reward becomes negative after an unsuccessful evaluation ($\bar{\rho}(t) < 0$). Also, parasites might replicate independently of their host if they accrue a reward: 
        \begin{eqnarray}
          \bar{\rho}(t) > 2 \bar{n}\bar{\rho}_0. 
        \end{eqnarray}
      In this case, the daughter parasite will occupy a new, random spot in the shadow ecosystem. By doing so, it might displace an earlier parasite---thus changing the parasite that previously infected some other host. Parasite replication is resolved before host death, so that the daughter of the parasite of a dying host (as in Fig.\ \ref{supFig:2}) lives on. 

      The model is limited in that guessers must be smaller than their environment. To ensure this, if a daughter parasite was relocated to a spot in which a host is less complex than the incoming parasite, the host is removed from the experiment altogether and the parasite is promoted to the main ecosystem. We understand that this is what happens if a parasite is relocated to an empty spot, where a $0$-guesser dwelt. Our experiments indicate that this mechanism is seldom used, since hosts and parasites sustain a gap of complexity that prevents parasites from becoming more complex than hosts. 

      The final element of this experiment is the evolutionary dynamics. Besides the mode of replication just described, both hosts and parasites might mutate. Mutation is enabled only if a guesser accrues enough reward to satisfy the metabolic load of a guesser with just a unit longer (i.e.\ $(n+1)\rho_0$ for hosts, $(\bar{n}+1)\bar{\rho}_0$ for parasites). This is, mutation is possible if $$\rho(t) > (2n+1)\rho_0$$ for hosts and if $$\bar{\rho}(t) > (2\bar{n}+1)\bar{\rho}_0$$ for parasites. If this happens, mutation ensues with probability $p_{\mu}$. If a mutation takes place, it results in a guesser of less complexity ($n \rightarrow n-1$ for hosts, $\bar{n} \rightarrow \bar{n}-1$ for parasites) or in a guesser of more complexity ($n \rightarrow n-1$ for hosts, $\bar{n} \rightarrow \bar{n}-1$ for parasites) with equal chance.

\end{document}